# Third order perturbed modified Heisenberg Hamiltonian of fcc structured ferromagnetic films with seventy spin layers

P. Samarasekara [1, 2, 4], N.U.S. Yapa [2, 3] and C.A.N. Fernando [4]

[1]Department of Physics, University of Peradeniya, Peradeniya, Sri Lanka

[2]Postgraduate Institute of science, University of Peradeniya, Sri Lanka

[3]Department of Physics, Open University of Sri Lanka, Kandy, Sri Lanka

[4] Department of Nano Science Technology, Wayamba University of Sri Lanka, Kuliyapitiya, Sri Lanka

**Abstract:**
Magnetic properties of fcc structured ferromagnetic films with the number of spin layers up to seventy was described using third order perturbed Heisenberg Hamiltonian. The variation of magnetic easy direction, magnetic energies in easy and hard directions, magnetic anisotropy energy and the angle between easy and hard directions was investigated by varying the number of spin layers. Spin exchange interaction, magnetic dipole interaction, second and fourth order magnetic anisotropies, in and out of plane applied magnetic fields, demagnetization factor and stress induced anisotropy were considered in the model. Because magnetic dipole interaction and demagnetization factor represent microscopic and macroscopic properties of the sample, respectively, both these terms were incorporated in our theoretical model. Although our model is a semi-classical model, some discrete variations of angle of easy axis were observed. Our theoretical data qualitatively agree with experimental data of Fe and Ni ferromagnetic films.

**1. Introduction:**
Ferromagnetic films find potential applications in magnetic memory devices and microwave devices. Magnetic thin films are employed in miniature magnetic devices. Magnetic easy axis oriented films provide the same magnetic properties as bulk magnetic materials. Energy density of magnetic easy axis oriented films is almost same as that of bulk magnetic materials. However, the detailed theoretical studies related to the easy axis orientation are limited. The quasistatic magnetic hysteresis of ferromagnetic thin films grown on a vicinal substrate has been theoretically explored using Monte Carlo simulations [1]. Structural and magnetic properties of two dimensional FeCo ordered alloys have been investigated by first principles band structure theory [2]. EuTe films with surface elastic stresses have been theoretically studied using Heisenberg Hamiltonian [3]. De Vries theory was employed to explain the magnetostriction of dc magnetron sputtered FeTaN thin films [4]. Magnetic layers of Ni on Cu have been theoretically investigated using the Korringa-Kohn-Rostoker Green's function method [5]. Electric and magnetic properties of multiferroic thin films have been theoretically described using modified Heisenberg model and transverse Ising model coupled with Green's function technique [6].

Previously ferromagnetic films with four and five spin layers have been described using second order perturbed Heisenberg Hamiltonian by us [7, 8]. Thick ferromagnetic films have been explained using second and third order perturbed Heisenberg Hamiltonian [9, 10]. In addition, ferromagnetic films with three spin layers have been studied using third order perturbed Heisenberg Hamiltonian [11]. The interfacial coupling dependence of the magnetic ordering in ferro-antiferromagntic bilayers has been studied using the Heisenberg Hamiltonian [12]. Heisenberg Hamiltonian incorporated with spin exchange interaction, magnetic dipole interaction, applied magnetic field, second and fourth order magnetic anisotropy terms has been solved for ferromagnetic thin films [13, 14, 15]. The domain structure and Magnetization reversal in thin magnetic films was described using computer simulations [16]. Heisenberg Hamiltonian has been employed to theoretically describe in-plane dipole coupling anisotropy of a square ferromagnetic Heisenberg monolayer [17].

Variation of magnetic easy axis of fcc structured ferromagnetic films with number of spin layers is described in this manuscript. MATLAB computer program was applied to plot all the 3-D and 2-D graphs.

## 2. Model:

The classical Heisenberg Hamiltonian of ferromagnetic thin films can be given in following form.

$$H = -\frac{J}{2}\sum_{m,n}\vec{S}_m \cdot \vec{S}_n + \frac{\omega}{2}\sum_{m \neq n}\left(\frac{\vec{S}_m \cdot \vec{S}_n}{r_{mn}^3} - \frac{3(\vec{S}_m \cdot \vec{r}_{mn})(\vec{r}_{mn} \cdot \vec{S}_n)}{r_{mn}^5}\right) - \sum_m D_{\lambda_m}^{(2)}(S_m^z)^2 - \sum_m D_{\lambda_m}^{(4)}(S_m^z)^4$$

$$-\sum_{m,n}[\vec{H} - (N_d \vec{S}_n / \mu_0)] \cdot \vec{S}_m - \sum_m K_s \sin 2\theta_m$$

Above equation will be deduced to following form [13, 14, 15]

$$E(\theta) = -\frac{1}{2}\sum_{m,n=1}^{N}\left[(JZ_{|m-n|} - \frac{\omega}{4}\Phi_{|m-n|})\cos(\theta_m - \theta_n) - \frac{3\omega}{4}\Phi_{|m-n|}\cos(\theta_m + \theta_n)\right]$$

$$-\sum_{m=1}^{N}(D_m^{(2)}\cos^2\theta_m + D_m^{(4)}\cos^4\theta_m + H_{in}\sin\theta_m + H_{out}\cos\theta_m)$$

$$+\sum_{m,n=1}^{N}\frac{N_d}{\mu_0}\cos(\theta_m - \theta_n) - K_s\sum_{m=1}^{N}\sin 2\theta_m \qquad (1)$$

Here m and n represent indices of two different layers, N is the number of layers measured in direction perpendicular to the film plane, J is the magnetic spin exchange interaction, $Z_{|m-n|}$ is the number of nearest spin neighbors, $\omega$ is the strength of long range dipole interaction, $\Phi_{|m-n|}$ are constants for partial summation of dipole interaction, $D_m^{(2)}$ and $D_m^{(4)}$ are second and fourth order anisotropy constants, $H_{in}$ and $H_{out}$ are components of applied magnetic field, $N_d$ is the demagnetization factor, and $K_s$ is the constant related to the stress which depends on the magnetization and the magnitude of stress.

For non-oriented films, above angles $\theta_m$ and $\theta_n$ measured with film normal can be expressed in forms of $\theta_m = \theta + \varepsilon_m$ and $\theta_n = \theta + \varepsilon_n$, and above energy can be expanded up to the third order of ε as following,

$$E(\theta)=E_0+E(\varepsilon)+E(\varepsilon^2)+E(\varepsilon^3) \tag{2}$$

Here $E_0 = -\frac{1}{2}\sum_{m,n=1}^{N}(JZ_{|m-n|} - \frac{\omega}{4}\Phi_{|m-n|}) + \frac{3\omega}{8}\cos 2\theta \sum_{m,n=1}^{N}\Phi_{|m-n|}$

$$-\cos^2\theta\sum_{m=1}^{N}D_m^{(2)} - \cos^4\theta\sum_{m=1}^{N}D_m^{(4)} - N(H_{in}\sin\theta + H_{out}\cos\theta - \frac{N_d}{\mu_0} + K_s\sin 2\theta) \tag{3}$$

$E(\varepsilon) = -\frac{3\omega}{8}\sin 2\theta \sum_{m,n=1}^{N}\Phi_{|m-n|}(\varepsilon_m + \varepsilon_n) + \sin 2\theta\sum_{m=1}^{N}D_m^{(2)}\varepsilon_m + 2\cos^2\theta\sin 2\theta\sum_{m=1}^{N}D_m^{(4)}\varepsilon_m$

$-H_{in}\cos\theta\sum_{m=1}^{N}\varepsilon_m + H_{out}\sin\theta\sum_{m=1}^{N}\varepsilon_m - 2K_s\cos 2\theta\sum_{m=1}^{N}\varepsilon_m$

$E(\varepsilon^2) = \frac{1}{4}\sum_{m,n=1}^{N}(JZ_{|m-n|} - \frac{\omega}{4}\Phi_{|m-n|})(\varepsilon_m - \varepsilon_n)^2 - \frac{3\omega}{16}\cos 2\theta\sum_{m,n=1}^{N}\Phi_{|m-n|}(\varepsilon_m + \varepsilon_n)^2$

$-(\sin^2\theta - \cos^2\theta)\sum_{m=1}^{N}D_m^{(2)}\varepsilon_m^2 + 2\cos^2\theta(\cos^2\theta - 3\sin^2\theta)\sum_{m=1}^{N}D_m^{(4)}\varepsilon_m^2$

$+\frac{H_{in}}{2}\sin\theta\sum_{m=1}^{N}\varepsilon_m^2 + \frac{H_{out}}{2}\cos\theta\sum_{m=1}^{N}\varepsilon_m^2 - \frac{N_d}{2\mu_0}\sum_{m,n=1}^{N}(\varepsilon_m - \varepsilon_n)^2$

$+2K_s\sin 2\theta\sum_{m=1}^{N}\varepsilon_m^2$

$$E(\varepsilon^3) = \frac{\omega}{16}\sin 2\theta \sum_{m,n=1}^{N}(\varepsilon_m + \varepsilon_n)^3 \phi_{|m-n|} - \frac{4}{3}\cos\theta\sin\theta\sum_{m=1}^{N}D_m^{(2)}\varepsilon_m^3$$

$$-4\cos\theta\sin\theta(\frac{5}{3}\cos^2\theta - \sin^2\theta)\sum_{m=1}^{N}D_m^{(4)}\varepsilon_m^3 + \frac{H_{in}}{6}\cos\theta\sum_{m=1}^{N}\varepsilon_m^3$$

$$-\frac{H_{out}}{6}\sin\theta\sum_{m=1}^{N}\varepsilon_m^3 + \frac{4K_s}{3}\cos 2\theta\sum_{m=1}^{N}\varepsilon_m^3$$

After using the constraint $\sum_{m=1}^{N}\varepsilon_m = 0$, $E(\varepsilon) = \vec{\alpha}.\vec{\varepsilon}$

Here $\vec{\alpha}(\varepsilon) = \vec{B}(\theta)\sin 2\theta$ are the terms of matrices with

$$B_\lambda(\theta) = -\frac{3\omega}{4}\sum_{m=1}^{N}\Phi_{|\lambda-m|} + D_\lambda^{(2)} + 2D_\lambda^{(4)}\cos^2\theta \qquad (4)$$

Also $E(\varepsilon^2) = \frac{1}{2}\vec{\varepsilon}.C.\vec{\varepsilon}$

Here the elements of matrix C can be given as following,

$$C_{mn} = -(JZ_{|m-n|} - \frac{\omega}{4}\Phi_{|m-n|}) - \frac{3\omega}{4}\cos 2\theta \Phi_{|m-n|} + \frac{2N_d}{\mu_0}$$

$$+\delta_{mn}\{\sum_{\lambda=1}^{N}[JZ_{|m-\lambda|} - \Phi_{|m-\lambda|}(\frac{\omega}{4} + \frac{3\omega}{4}\cos 2\theta)] - 2(\sin^2\theta - \cos^2\theta)D_m^{(2)}$$

$$+4\cos^2\theta(\cos^2\theta - 3\sin^2\theta)D_m^{(4)} + H_{in}\sin\theta + H_{out}\cos\theta - \frac{4N_d}{\mu_0} + 4K_s\sin 2\theta\} \qquad (5)$$

Also $E(\varepsilon^3) = \varepsilon^2 \beta.\vec{\varepsilon}$

Here matrix elements of matrix β can be given as following,

$$\beta_{mn} = \frac{3\omega}{8}\sin 2\theta \Phi_{|m-n|} + \delta_{mn}\{\frac{\omega}{8}\sin 2\theta[A_m - \Phi_0] - \frac{4}{3}\cos\theta\sin\theta D_m^{(2)}$$

$$- 4\cos\theta\sin\theta(\frac{5}{3}\cos^2\theta - \sin^2\theta)D_m^{(4)} + \frac{H_{in}}{6}\cos\theta - \frac{H_{out}}{6}\sin\theta$$

$$+ \frac{4K_s}{3}\cos 2\theta\} \tag{6}$$

Also $\beta_{nm}=\beta_{mn}$ and, matrix β is symmetric.

Here $A_m$ values are different for even and odd N values, and can be given as following.

For odd N, $A_{\frac{N}{2}+0.5+n} = 2\sum_{v=0}^{\frac{N}{2}-0.5-n}\Phi_v + \sum_{v=\frac{N}{2}+0.5-n}^{\frac{N}{2}+n-0.5}\Phi_v$ for $m > \frac{N}{2}$,

where n varies from 1 to $\frac{N}{2} - 0.5$.

When n=0, $A_{\frac{N}{2}+0.5+n} = 2\sum_{v=0}^{\frac{N}{2}-0.5-n}\Phi_v$

$A_m$ for $m < \frac{N}{2}$ can be obtained using $A_{\frac{N}{2}+0.5+n} = A_{\frac{N}{2}+0.5-n}$

For even N, $A_{\frac{N}{2}+1+n} = 2\sum_{v=0}^{\frac{N}{2}-1-n}\Phi_v + \sum_{v=\frac{N}{2}-n}^{\frac{N}{2}+n}\Phi_v$ for $m > \frac{N}{2}$,

where n varies from 0 to $\frac{N}{2}-1$.

$A_m$ for $m < \frac{N}{2}$ can be obtained using $A_{\frac{N}{2}+1+n} = A_{\frac{N}{2}-n}$

Therefore, the total magnetic energy given in equation 2 can be deduced to

$$E(\theta)=E_0+ \vec{\alpha}.\vec{\varepsilon} +\frac{1}{2}\vec{\varepsilon}.C.\vec{\varepsilon} + \varepsilon^2 \vec{\beta}.\vec{\varepsilon} \qquad (7)$$

Because the derivation of a final equation for ε with the third order of ε in above equation is tedious, only the second order of ε will be considered for following derivation.

Then $E(\theta)=E_0+ \vec{\alpha}.\vec{\varepsilon} +\frac{1}{2}\vec{\varepsilon}.C.\vec{\varepsilon}$

Using a suitable constraint in above equation, it is possible to show that $\vec{\varepsilon} = -C^+.\vec{\alpha}$

Here $C^+$ is the pseudo-inverse given by

$$C.C^+ = 1-\frac{E}{N}. \qquad (8)$$

E is the matrix with all elements given by $E_{mn}=1$.

After using ε in equation 7, energy per unit spin is given by

$$E(\theta)=E_0 -\frac{1}{2}\vec{\alpha}.C^+.\vec{\alpha} - (C^+\alpha)^2 \vec{\beta}(C^+\alpha) \qquad (9)$$

**3. Results and discussion:**

For complex lattices, magnetic dipole interactions were calculated by considering the interaction between each pair of spins in lattice [9, 18]. When two different types of atoms occupy different lattice sites, the spin contribution from each different type of atoms would be considered. However, for lattice of fcc(001) lattice, $Z_0=4$, $Z_1=4$, $Z_2=0$ and $\Phi_0=9.0336$, $\Phi_1=1.4294$ [13, 14, 15]. According to our experimental studies, the magnetic anisotropy of magnetic thin films depend on stress induced anisotropy [19, 20, 21]. Stress of magnetic films arises in the heating or cooling process due to the difference between thermal expansion coefficients of substrate and film. Figure 1 indicates the 3-D plot of total magnetic energy per unit spin versus angle and stress induced anisotropy for film with 30 spin layers (N=30). Other energy parameters were kept at $\frac{J}{\omega}=\frac{H_{in}}{\omega}=\frac{N_d}{\mu_0\omega}=\frac{H_{out}}{\omega}=10$, $\frac{D_m^{(2)}}{\omega}=30$ and $\frac{D_m^{(4)}}{\omega}=20$. One minimum and a maximum of this plot can be observed at $\frac{K_s}{\omega}=18$ and $\frac{K_s}{\omega}=31$, respectively. This 3-D plot is entirely different from the same 3-D plot of five layered fcc lattice obtained using second order perturbed Heisenberg Hamiltonian [8]. Although the total magnetic energy ($\frac{E(\theta)}{\omega}$) of five layered films varies up to 600, total energy of film with 30 spin layers varies up to $10^6$. Reason is attributed to the fact that the total magnetic energy increases with the total number of spins in the film. Compared to the 3-D plot given in this manuscript, the peaks of the same 3-D plot of fcc structured ferromagnetic films with three layers obtained using third order perturbed Heisenberg Hamiltonian are widely separated [11]. Furthermore, $\frac{E(\theta)}{\omega}$ with three layered films changes up to $10^4$. However, the shape of the 3-D plot of film with 30 spin layers is fairly close to the shape of 3-D plot of three layered film.

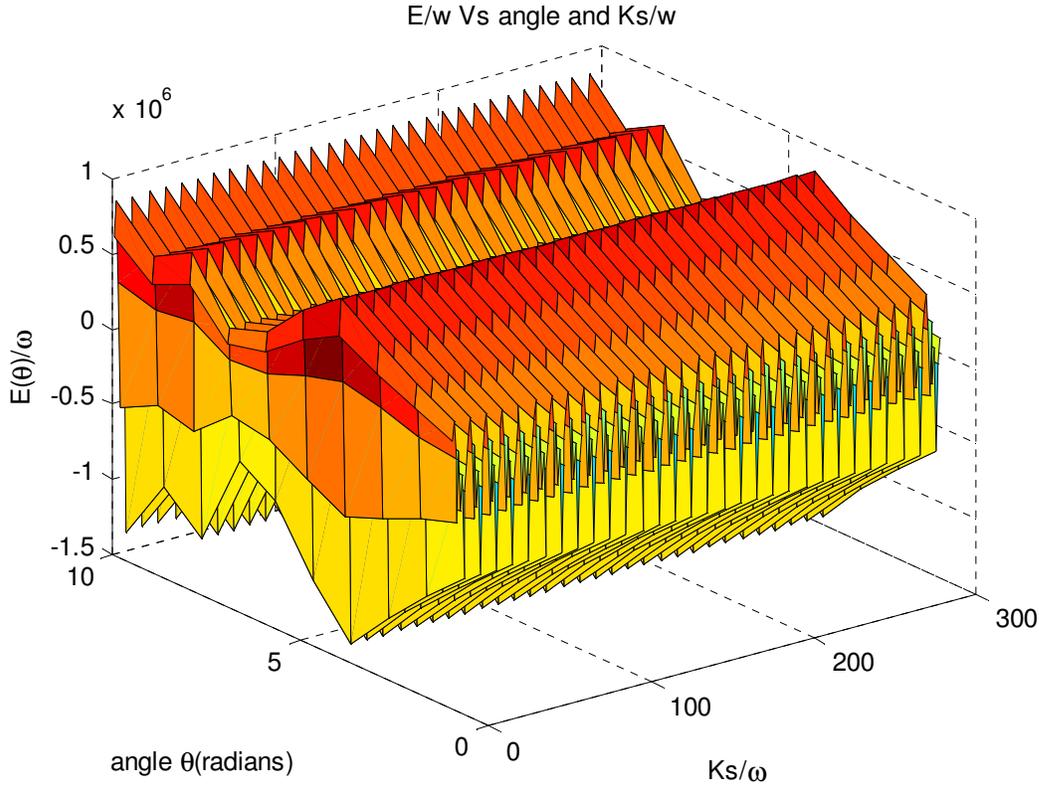

Figure 1: 3-D plot of $\dfrac{E(\theta)}{\omega}$ versus azimuthal angle and $\dfrac{K_s}{\omega}$.

One minimum of above 3-D plot was observed at $\dfrac{K_s}{\omega}=18$. Graph of energy versus angle at $\dfrac{K_s}{\omega}=18$ is given in figure 2. A minimum and a maximum of this graph can be observed at $18.0023^0$ and $122.3838^0$, respectively. Therefore, the magnetic easy direction is $18.0023^0$. One maximum of above 3-D plot was observed at $\dfrac{K_s}{\omega}=31$. Figure 3 shows the graph of energy versus angle at $\frac{K_s}{\omega}$ =31. A minimum and a maximum of this graph can be observed at $25.1987^0$ and $125.9935^0$, respectively. Magnetic hard direction is $125.9935^0$. Therefore, the angle between magnetic easy and hard directions is $107.9912^0$.

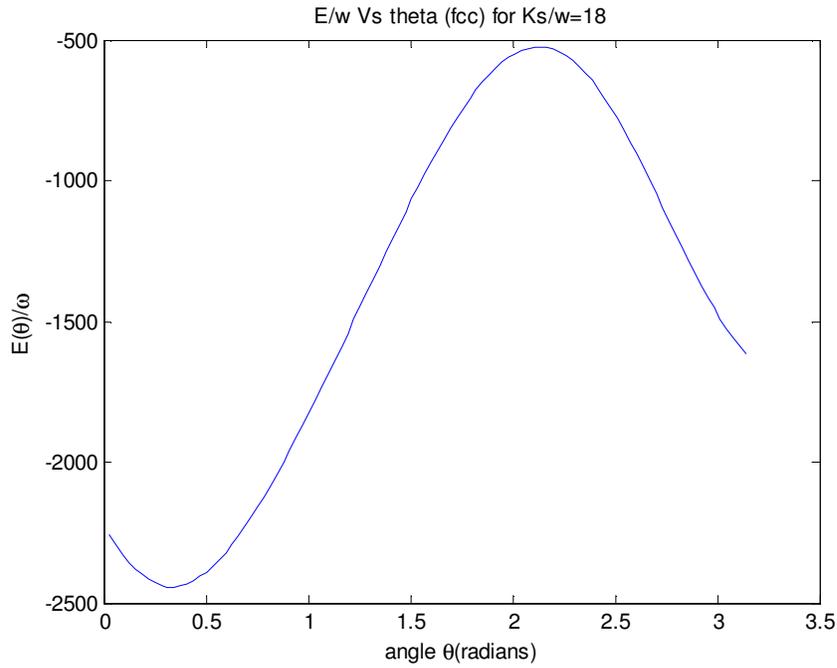

Figure 2: Graph of $\frac{E(\theta)}{\omega}$ versus azimuthal angle for $\frac{K_s}{\omega}$ =18.

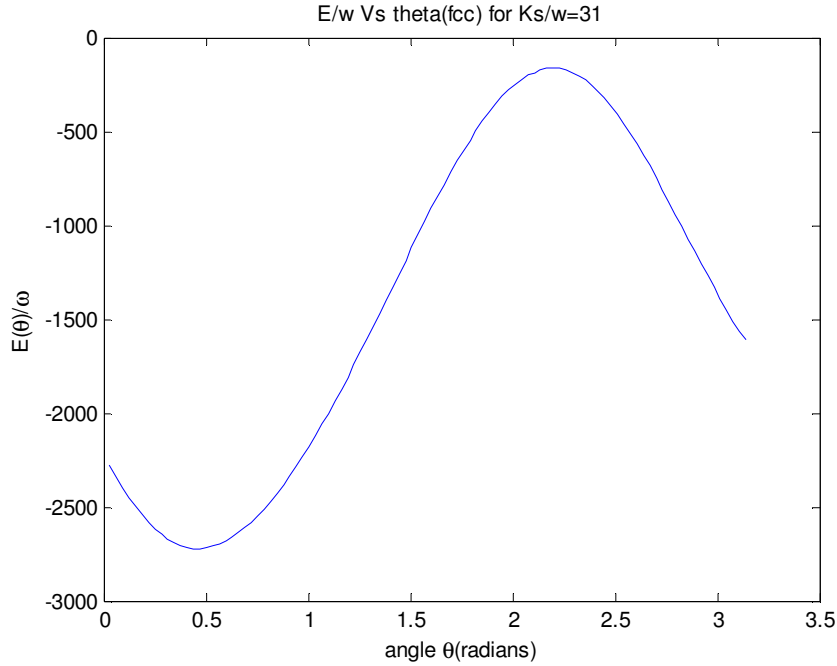

Figure 3: Graph of $\frac{E(\theta)}{\omega}$ versus azimuthal angle for $\frac{K_s}{\omega}=31$.

Figure 4 shows the 3-D plot of energy versus angle and $\frac{J}{\omega}$ for N=30. Other parameters were kept at $\frac{H_{in}}{\omega}=\frac{K_s}{\omega}=\frac{H_{out}}{\omega}=\frac{N_d}{\mu_0\omega}=10$, $\frac{D_m^{(2)}}{\omega}=30$ and $\frac{D_m^{(4)}}{\omega}=20$. One minimum and a maximum of this 3-D plot were observed at $\frac{J}{\omega}=10$ and $\frac{J}{\omega}=1$, respectively. After plotting graphs of energy versus angle for $\frac{J}{\omega}=10$, a minimum and a maximum were observed at 12.5993⁰ and 122.3838⁰, respectively. Similarly by plotting graphs of energy versus angle for $\frac{J}{\omega}=1$, a minimum and a maximum were observed at 70.1874⁰ and 122.3838⁰, respectively. Therefore, magnetic easy and hard directions were found to be 12.5993⁰ and 122.3838⁰, respectively. Angle between magnetic easy and hard directions is 109.7845⁰. Peaks in the same 3-D plot of fcc structured ferromagnetic films with three layers obtained using third order perturbed Heisenberg Hamiltonian were well separated compared to this 3-D plot for N=30 [11]. Energy ($\frac{E(\theta)}{\omega}$) of film with 3 spin layers varies up to 200, while the energy of film with 30 spin layers varies up to 5000. Because total number of spins in the film increases with the number of spin layers, the total magnetic energy increases with the number of spin layers.

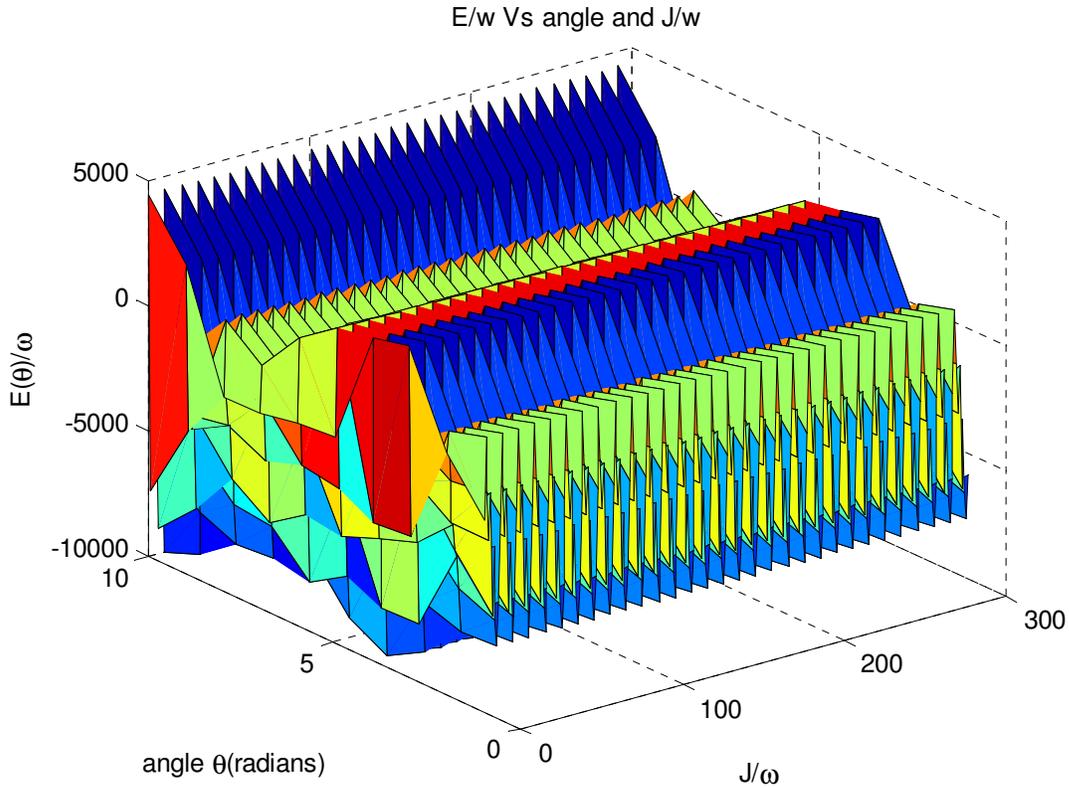

Figure 4: 3-D plot of $\frac{E(\theta)}{\omega}$ versus azimuthal angle and $\frac{J}{\omega}$.

Table 1 shows the variation of easy directions and corresponding energies with the number of spin layers. All other parameters were kept at $\frac{H_{in}}{\omega} = \frac{J}{\omega} = \frac{K_s}{\omega} = \frac{H_{out}}{\omega} = \frac{N_d}{\mu_0 \omega} = 10$, $\frac{D_m^{(2)}}{\omega} = 30$ and $\frac{D_m^{(4)}}{\omega} = 20$. The magnetic easy direction discretely rotates from out of plane to in plane direction of the film with increase of the number of spin layers. The total magnetic energy in either easy or hard directions gradually increases with the number of spin layers. In addition, the energy required to rotate spins from easy to hard directions (magnetic anisotropy energy) gradually increases with the number of spin layers. Furthermore, angle between easy and hard directions slightly varies with the number of spin layers. The same variations could be observed for fcc structured ferromagnetic films using second order perturbed Heisenberg Hamiltonian [22].

| | θ(easy) In degrees | E/ω (easy) | E/ω (hard) | ΔE= E(easy)-E(hard) | Δθ=θ(hard)-θ(easy) In degrees |
|---|---|---|---|---|---|
| **10** | 16.5986 | -678.2 | -230.3 | 447.9 | 104.0091 |
| **20** | 30.6017 | -1031 | -319.1 | 711.9 | 106.2207 |
| **30** | 30.6017 | -1533 | -484.6 | 1048.4 | 106.2207 |
| **40** | 30.6017 | -2074 | -649.3 | 1424.7 | 107.9968 |
| **50** | 30.6017 | -2596 | -815.6 | 1780.4 | 106.2207 |
| **60** | 30.6017 | -3118 | -980.0 | 2138.0 | 107.9968 |
| **70** | 32.4008 | -3638 | -1145 | 2493 | 106.1977 |

Table 1: Magnetic properties of fcc structured ferromagnetic films.

Magnetic easy direction experimentally depends on substrate temperature, thickness of the films, orientation of substrate, deposition rate, distance between target and substrate, type of sputtering gas and sputtering pressure [21]. The variation of magnetic easy axis with temperature has been theoretically explained using Heisenberg Hamiltonian by us [23, 24, 25]. Spin reorientation and the rotation of one spin component in 2-D model were investigated in that theoretical study.
Magnetic easy axis of Ni and Fe ferromagnetic films experimentally rotates from out of plane to in plane direction [26, 27]. Therefore, our theoretical data qualitatively agree with experimental data. Quantitative analysis can be performed, only if the experimental values of magnetic parameters such as J, ω, $H_{in}$, $H_{out}$, $N_d$, $K_s$, $D_m^{(2)}$ and $D_m^{(4)}$ are available. Because the experimental values of these magnetic parameters vary from sample to sample, it is difficult to perform any quantitative analysis.

## 4. Conclusion:

As the number of spin layers was increased from 10 to 70, the magnetic easy axis rotates from out of plane to in plane direction. However, the rotation is not smooth. Sudden changes can be observed from 10 to 20 and 60 to 70 spin layers. But the easy axis does not vary in the region from 20 to 60 spin layers. This indicates some discrete or quantum mechanical behavior of this model, although our model is a semi-classical model. When the final equation of total magnetic energy was found using equation number 9, there were some terms with N (number of spin layers) in the denominator of the final equation of the energy. This can be the reason for this abnormal variation of the angle of easy axis. In addition, energy in either easy or hard directions gradually increases with the number of spin layers, because the total number of spins in the film gradually increases with the number of spin layers. The magnetic anisotropy energy also gradually increases with the number of spin layers. However, the angle between easy and hard directions doesn't change by a considerable amount systematically. Our theoretical data qualitatively agree with experimental data of Fe and Ni ferromagnetic films.

**References:**

1. D. Zhao, Feng Liu, D.L. Huber and M.G. Lagally (2002). Step-induced magnetic-hysteresis anisotropy in ferromagnetic thin films. *Journal of Applied Physics* **91(5)**: 3150-3153.

2. D. Spisak and J. Hafner. (2005). Theoretical study of FeCo/W(110) surface alloys. *Journal of Magnetism and Magnetic Materials* **286**: 386-389.

3. Radomska Anna and Balcerzak Tadeusz (2003). Theoretical studies of model thin EuTe films with surface elastic stresses. *Central European Journal of Physics* **1(1)**: 100-117.

4. C. Cates James and Alexander Jr Chester. (1994). Theoretical study of magnetostriction in FeTaN thin films. *Journal of Applied Physics* **75**: 6754-6756.

5. A. Ernst, M. Lueders, W.M. Temmerman, Z. Szotek and G. Van der Laan (2000). Theoretical study of magnetic layers of nickel on copper; dead or live?. *Journal of Physics: Condensed matter* **12(26)**: 5599-5606.

6. Kovachev St and J.M. Wesselinowa (2009). Theoretical study of multiferroic thin films based on a microscopic model. *Journal of Physics: Condensed matter* **21(22)**: 225007.

7. P. Samarasekara (2008). Four layered ferromagnetic ultra-thin films explained by second order perturbed Heisenberg Hamiltonian. *Ceylon Journal of Science* **14**: 11-19.

8. P. Samarasekara and B.I. Warnakulasooriya (2016). Five layered fcc ferromagnetic films as described by modified second order perturbed Heisenberg Hamiltonian. *Journal of science: University of Kelaniya* **11**: 11-21.

9. P. Samarasekara and Amila D. Ariyaratne (2012). Determination of magnetic properties of Cobalt films using second order perturbed Heisenberg Hamiltonian. *Research & Reviews: Journal of Physics-STM journals* **1(1)**: 16-23.

10. P. Samarasekara (2008). Influence of third order perturbation on Heisenberg Hamiltonian of thick ferromagnetic films. *Electronic Journal of Theoretical Physics* **5(17):** 227-236.

11. P. Samarasekara and N.U.S. Yapa (2016). Third order perturbed energy of fcc ferromagnetic thin films as described by Heisenberg Hamiltonian. *Ceylon Journal of Science* **45(2)**: 71-77.

12. Tsai Shan-Ho, D.P. Landau and C. Schulthess Thomas (2003). Effect of interfacial coupling on the magnetic ordering in ferro-antiferromagntic bilayers. *Journal of Applied Physics* **93(10)**: 8612-8614.

13. A. Hucht and K.D. Usadel. (1997). Reorientation transition of ultrathin ferromagnetic films. *Physical Review B* **55**: 12309.

14. A. Hucht and K.D. Usadel (1999). Theory of the spin reorientation transition of ultra-thin ferromagnetic films. *Journal of Magnetism and Magnetic materials* **203(1)**: 88-90.

15. K.D. Usadel and A. Hucht (2002). Anisotropy of ultrathin ferromagnetic films and the spin reorientation transition. *Physical Review B* **66**: 024419.

16. U. Nowak (1995). Magnetisation reversal and domain structure in thin magnetic films: Theory and computer simulation. *IEEE transaction on magnetics* **31(6-2)**: 4169-4171.

17. M. Dantziger, B. Glinsmann, S. Scheffler, B. Zimmermann and P.J. Jensen (2002). In-plane dipole coupling anisotropy of a square ferromagnetic Heisenberg monolayer. *Physical Review B* **66**: 094416.

18. P. Samarasekara (2014). Cobalt ferrite films as described by third order perturbed Heisenberg Hamiltonian. *Journal of science: University of Kelaniya* **9**: 27-38.

19. P. Samarasekara and F.J. Cadieu (2001). Polycrystalline Ni ferrite films deposited by RF sputtering techniques. *Japanese Journal of Applied Physics* **40:** 3176-3179.

20. P. Samarasekara and F.J. Cadieu (2001). Magnetic and Structural Properties of RF Sputtered Polycrystalline Lithium Mixed Ferrimagnetic Films. *Chinese Journal of Physics* **39(6):** 635-640.

21. P. Samarasekara (2002). Easy Axis Oriented Lithium Mixed Ferrite Films Deposited by the PLD Method. *Chinese Journal of Physics* **40(6):** 631-636.

22. P. Samarasekara and T.H.Y.I.K. De Silva (2017). Fcc and bcc structured ferromagnetic films with five to twenty spin layers explained by second order perturbed Heisenberg Hamiltonian. *Ceylon Journal of Science* **46(1)**: 37-45.

23. P. Samarasekara and N.H.P.M. Gunawardhane (2011). Explanation of easy axis orientation of ferromagnetic films using Heisenberg Hamiltonian. *Georgian electronic scientific journals: Physics* **2(6)**: 62-69.

24. P. Samarasekara and Udara Saparamadu (2012). Investigation of Spin Reorientation in Nickel Ferrite Films. *Georgian electronic scientific journals: Physics* **1(7):** 15-20.

25. P. Samarasekara and Udara Saparamadu (2013). In plane oriented Strontium ferrite thin films described by spin reorientation. *Research & Reviews: Journal of Physics-STM journals* **2(2)**: 12-16.

26. J. Araya-Pochet, C.A. Ballentine and J.L. Erskine (1988). Thickness and temperature dependent spin anisotropy of ultrathin epitaxial Fe films on Ag(100). *Physical Review B* **38(11)**: 7846-7849.

27. U. Parlak, M.E. Akoz, S. Tokdemir Ozturk and M. Erkovan (2015). Thickness dependent magnetic properties of polycrystalline nickel thin films. *Acta Physica Polonica A* **127(4)**: 995-997.18